# Analysis of social media content and search behavior related to seasonal topics using the sociophysics approach

**Akira Ishii, Toshimichi Wakabayashi, Nozomi Okano**
Department of Applied Mathematics and Physics, Tottori University
Koyama, Tottori, 680-8552, Japan

**Yasuko Kawahata**
Faculty of Social and Information Studies, 4-2 Aramaki-machi
Maebashi, Gunma, 371-8510, Japan

## ABSTRACT

We studied the time interval between posting social media content and search action related to seasonal topics. The analysis was performed using a mathematical model of the search behavior as in the theory of sociophysics. As seasonal topics, the word "cherry blossom" was considered for spring, "bikini" for summer, "autumn leaves" for fall, and "skiing" for winter. We examined the influence of blogs and Twitter posts given the search behavior and found a time deviation of interest on these topics.

**Keywords**: Search behavior, Google Trend, mathematical model for hit phenomena, seasonal topics.

## 1. INTRODUCTION

The fluctuation of the interests of people in a society at different times of the year can be determined based on the content written on social media and the number of posts on each topic. However, there are many people who search for topics without writing on social media. Posts written on social media are on topics which are of obvious interest to people, whereas Internet search can be regarded as a potential interest attention of people. Thus, nowadays, search actions on the Internet correspond to intention of people in a society.

In 2005, Ettredge et al. [1] and Cooper et al. [2] suggested that Web search data was useful for forecasting social phenomena. There are many studies that use search action data as typical data to determine intentions. Investigation of the effects of mass media and social media on search actions is very interesting and meaningful. Although it is possible to investigate this data in time series using regression analysis, the sociophysics approach performs better in analyzing time series data.

In general, search actions are affected by mass media information, personal communications, and surrounding atmosphere in the society. Moreover, information on social media like Twitter, blogs, Facebook, Instagram, and YouTube encourages people to conduct searches on the Internet.

Recently, Ishii et al. presented a new theory [3] based on their mathematical model for hit phenomenon [4,5], where the number of daily posts on Twitter or blogs can be calculated using the sociophysics equation under the influence of mass media and its own social media. In the new mathematical model, the number of search actions are calculated and the effect of blog posts and Twitter posts on the search behavior related to the same topic can be analyzed.

In this study, we focus on social media content and search behavior related to seasonal topics to find some interesting features of the time intervals between social media posting and search actions. As can be seen in Fig. 1, the timings of Twitter posts, blog posts, and search actions are different. The posting on blogs and Twitter correspond to the overt layer in a society. On the other hand, search behavior corresponds to a latent layer of the society. Here, we select four topics for each season: "cherry blossom" in Japan for spring, "bikini" for summer, "autumn leaves" for fall, and "skiing" for winter.

As can be seen in Fig. 2 and Fig. 3, the words "skiing" and "bikini" exhibit seasonal behavior because winter is high season for skiing and summer is high season for wearing bikinis.

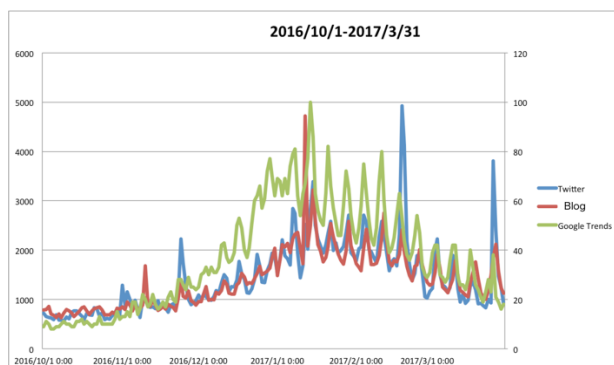

Fig. 1 Observed timings of Twitter posting, blog posting, and search action related to skiing. The values are arbitrarily normalized for comparison. The vertical value of the three values are arbitrarily normalized to fit into one graph.

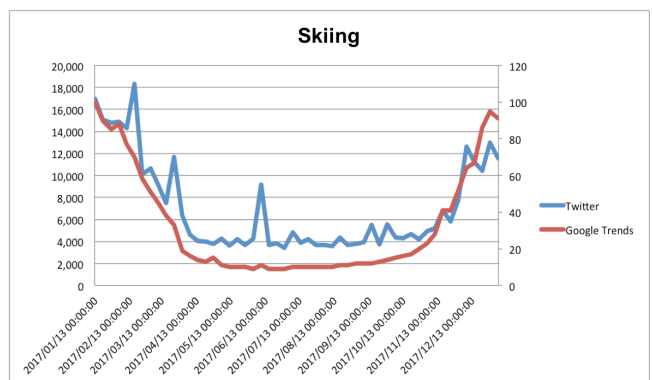

Fig. 2 Daily Twitter posting and search behavior related to skiing from January to December 2017.

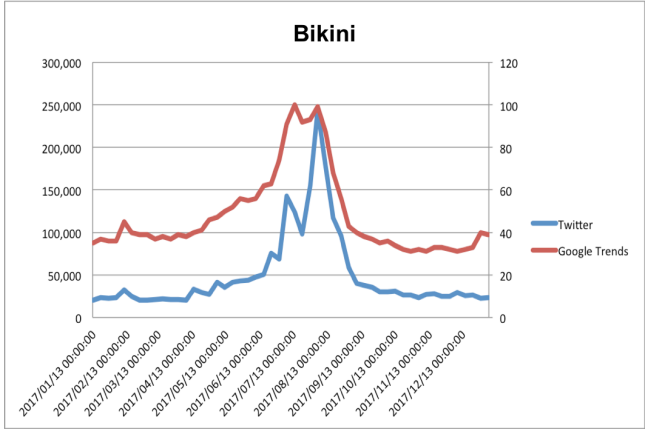

Fig. 3 Daily Twitter posting and search behavior related to bikini from January to December 2017.

## 2. THEORETICAL METHOD

In the theory of search behavior [3], the interest and concern on a certain topic can be calculated using a mathematical model of differential equations. Here, we introduce I(t) as the interest or concern on a certain topic. We construct a mathematical model based on the mathematical model for the hit phenomenon within a society presented as a stochastic process of interactions of human dynamics in the sense of many body theory in physics [4,5]. As in the model in [4,5], we assume that the intention of humans in a society is affected by the three factors: advertisement, communication with friends, and rumors. Advertisements act as external forces; communications with friends are a form of direct communication and its effect is considered as interaction with the intention of friends. The rumor effect is considered as the interaction among three persons and a form of indirect communication as described [1]. In the model, we use only the time distribution of advertisement budget as an input, and word-of-mouth (WOM) represented by posts on social network systems is the observed data for comparison with the calculated results. The parameters in the model are adjusted by the comparison with the calculated and observed social media posting data.

According to [4], we can write the equation for the intention of each person using the exponential form as follows:

$$\frac{dI_i(t)}{dt} = -aI_i(t) + \sum_j d_{ij}I_j(t) + \sum_j \sum_k h_{ijk}I_j(t)I_k(t) + f_i(t) \quad (1)$$

where $d_{ij}$, $h_{ijk}$, and $f_i(t)$ denote the coefficient of direct communication, coefficient of indirect communication, and random external force effect for person i, respectively. The random external force corresponds to advertisement via mass media. We consider the above equation for every consumer so that i = 1, …, $N_p$. Considering the effect of direct communication, indirect communication, and the decline of audience, we obtain the above equation for the mathematical model for hit phenomenon. The effect of advertisement and publicity for each person can be described as the mean field value of the random external force effect $<f_i(t)>$

Eq. (1) is for individual persons but it is not convenient to be used for analysis. Thus, we consider the ensemble average of the purchase intention of individual persons as follows:

$$\langle I(t) \rangle = \frac{1}{N}\sum_i I_i(t)$$

Taking the ensemble average of Eq. (1), we obtain the following form as the intention of a society as a collective mode,

$$\frac{d\langle I(t)\rangle}{dt} = -a\langle I(t)\rangle + D\langle I(t)\rangle + P\langle I(t)\rangle^2 + \langle f(t)\rangle \quad (2)$$

where $Nd = D$ and $N^2 p = P$. The detailed derivation is shown in [2].

In the new mathematical model for search behavior, we use daily blog and Twitter postings as the external force; therefore, we can write the equation as follows,

$$\frac{dI_i(t)}{dt} = -aI_i(t) + \sum_j d_{ij}I_j(t) + \sum_j \sum_k h_{ijk}I_j(t)I_k(t) + \sum_\xi C_\xi A_\xi(t) \quad (3)$$

where $\xi$ denotes the effect of TV, effect of news websites, effect of blogs, effect of Twitter, and effect of Wikipedia.

$$\frac{d\langle I(t)\rangle}{dt} = -a\langle I(t)\rangle + D\langle I(t)\rangle + P\langle I(t)\rangle^2 + \sum_\xi C_\xi A_\xi(t) \quad (4)$$

In the following calculation, coefficients $C_\xi$, $D$, $P$ are determined such that the calculated value according to Eqs. (2) or (4) coincides with the daily change of the observed data of blog posting, Twitter posting, and search action. The values can be obtained using the Monte Carlo method and the details can be found in [4].

The effects of advertisement and publicity are obtained from the dataset of M Data Co. Ltd. and WOM represented by posts on social network systems are observed using the system by Hottolink Co. Ltd. The search behavior is obtained using Google Trend.

For reliability of parameter determination, we introduce the "R-factor" (reliability factor), which is well-known in the field of low-energy electron diffraction (LEED) [6]. In LEED experiments, the experimentally observed curve of current vs. voltage is compared to the corresponding theoretical curve using the R-factor.

For our study, we define the R-factor as follows:

$$R = \frac{\sum_i (f(i) - g(i))^2}{\sum_i [f^2(i) + g^2(i)]} \quad (5)$$

where $f(i)$ and $g(i)$ correspond to the calculated $I(t)$ and the observed number of blog posts or tweets, respectively. The smaller the value of R, the better the functions $f$ and $g$. Thus, we use a random number to search for the parameter set that minimizes R. This random number technique is similar to the Metropolis method [7], which we used previously [4]. In actual calculations, we change each parameter within 10% of its value using the random number per turn. We perform such calculations for more than one-hundred-thousand turns. In our case, we try to obtain the parameter configuration with the minimum R-factor.

In actual calculations, for adjusting parameters $C_\xi$, $D$, and $P$, the local minimum trapping needs to be avoided. There are several ways to determine the minimum condition, including the steepest descent, equation of motion method, and conjugate gradient method. In this study, we only perform calculations using several initial values in a Metropolis-like method to avoid local minimum trapping. To check the accuracy of the parameter adjustment, we use the R-factor value. For every calculation shown in this study, the R-factor is below 0.01.

## 3. RESULTS AND DISCUSSION

We analyze search behavior for the word "bikini" using the new mathematical model in [3]. The observed value for search behavior is collected using Google Trend. The results are shown in Figs. 4-10. The calculations are done for the year 2017. The coefficients are calculated for each month and we assume that the coefficients have constant values in one month.

We calculate direct communication strength D, indirect communication strength P, the effect of advertisement on television $C_{adv\_t}$, the effect of news websites $C_{adv\_n}$, the effect of blogs $C_{adv\_blog}$, the effect of Twitter $C_{adv\_twitter}$ and the effect of Wikipedia $C_{wikipedia}$.

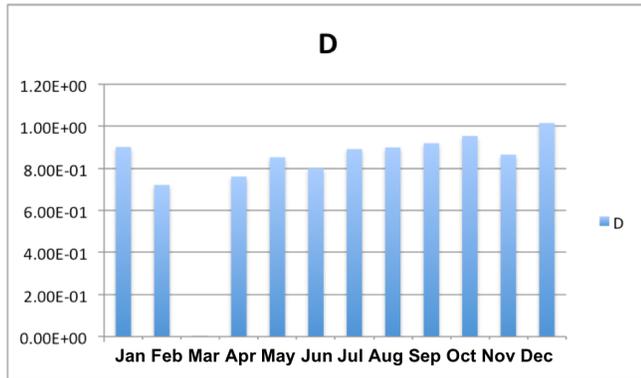

Fig. 4 Direct communication strength D for "bikini" from January to December 2017 for the whole world. Each coefficient is calculated every month.

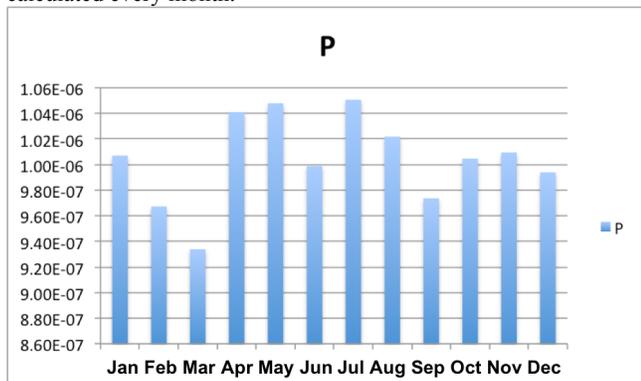

Fig. 5 Indirect communication strength P for "bikini" from January to December of 2017 for the whole world. Each coefficient is calculated every month.

In terms of the strength of direct communication D and indirect communication P, seasonality is not observed throughout the year. Moreover, for the effect of advertisement on television, seasonality is also not observed throughout the year. However, because the influence of news websites can be seen only in the summer, strong seasonality can be observed. Search is affected by blogs in June and search is affected by Twitter in April and May. Search actions influenced by Wikipedia are seen in March and April.

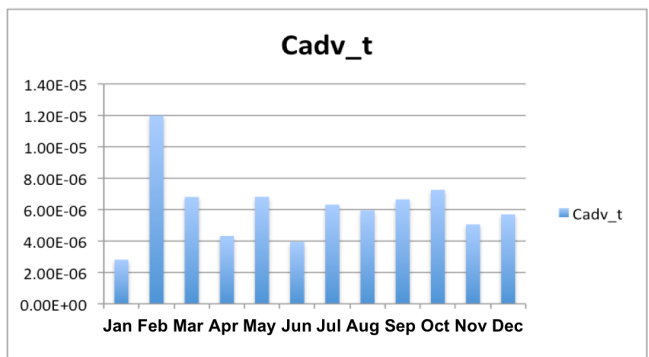

Fig. 6 Effect of advertisement on television for "bikini" from January to December 2017 for the whole world. Each coefficient is calculated every month.

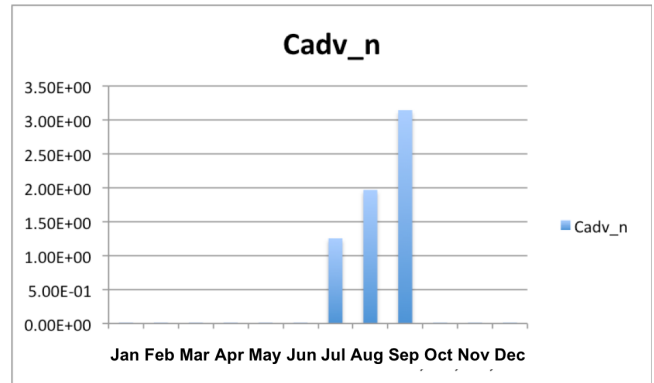

Fig. 7 Effect of news websites for "bikini" from January to December 2017 for the whole world. Each coefficient is calculated every month.

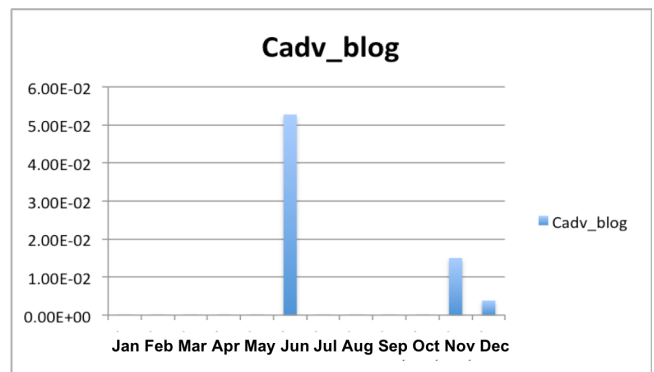

Fig. 8 Effect of blogs for "bikini" from January to December 2017 for the whole world. Each coefficient is calculated every month.

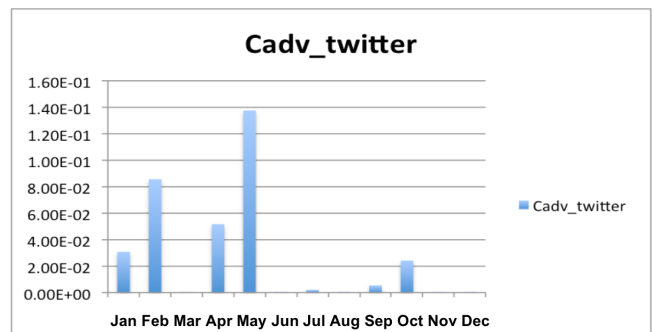

Fig. 9 Effect of Twitter for "bikini" from January to December 2017 for the whole world. Each coefficient is calculated every month.

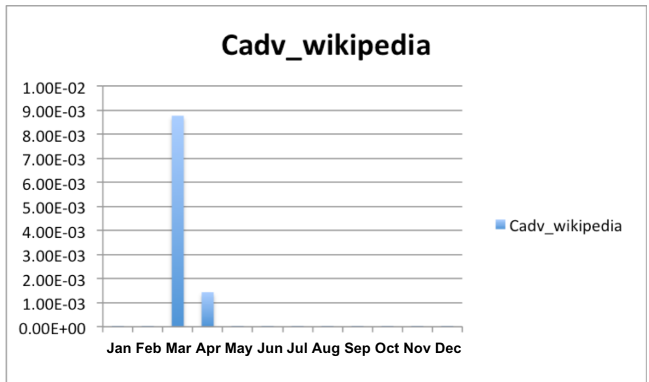

Fig. 10 Effect of Wikipedia for "bikini" from January to December 2017 for the whole world. Each coefficient is calculated every month.

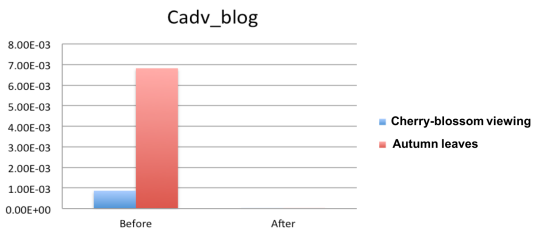

Fig. 11 Effect of blogs on search behavior before and after the best time for viewing cherry blossoms and autumn leaves in Japan.

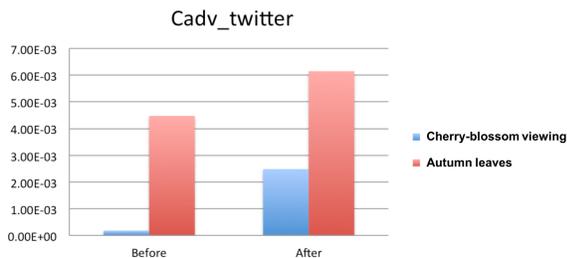

Fig. 11 Effect of Twitter on search behavior before and after the best time for viewing cherry blossoms and autumn leaves in Japan.

Because seasonal topics containing terms such as skiing, bikini, cherry blossom, autumn leaves, and so on are searched only during that season, this is a good example of observing the temporal transition of interest. As shown in Fig. 2 and Fig. 3, "skiing" is a topic of interest in winter only and "bikini" is of interest only in summer. Because the search results have been acquired worldwide in English, it can be seen that the seasonality of the Northern Hemisphere prevails even on a global scale.

Regarding the impact of "bikini" search, it can be seen from Figs. 8, 9, and 10 that the influence of blogs, Twitter, and Wikipedia on search behavior is strong before summer sea bathing season. This is because of the search action of women who plan to buy a new bikini for the sea bathing season.

In Figs. 11 and 12, for sightseeing of cherry blossoms and autumn leaves, which are spring and autumn events in Japan, respectively, blogs are used for search before the season and Twitter influences the search after the season ends. In other words, blogs are used for selecting the place for sightseeing of cherry blossoms and autumn leaves. In Japan, cherry blossom and autumn leaves can be seen only for a short time, so the location of sightseeing will be searched intensively during the seasons. Meanwhile, Twitter seemed to have been used to investigate what kind of cherry blossoms and autumn leaves were seen by other people after the seasons have ended. It is interesting to see the role sharing of Blog and Twitter as seen from the general public here.

## 4. CONCLUSIONS

We analyzed seasonal topics using a mathematical model of search behavior [3]. We studied the time interval between posting social media content and search behavior for seasonal topics in which the word "cherry blossom" was considered for spring, "bikini" for summer, "autumn leaves" for fall, "skiing" for winter. Using this new mathematical model, we checked the influence of blogs and Twitter given the search behavior and found a time deviation of interest on these topics. From this study, we found that the role of blog and Twitter in terms of search behavior is different for the people in a society.